\documentclass[5p,times]{elsarticle}
\usepackage{amsmath,amssymb,amsthm}
\usepackage{algorithm}
\usepackage{algorithmic}
\usepackage{graphicx}
\usepackage{booktabs}
\usepackage[section]{placeins}
\usepackage{multirow}
\usepackage[x11names,table]{xcolor}
\usepackage{tabularx}
\usepackage{gensymb}

\counterwithout{equation}{section}
\usepackage{makecell}
\usepackage{lineno}

\usepackage[colorlinks=true,           
            linkcolor=blue,            
            citecolor=blue,            
            urlcolor=blue]{hyperref}
\usepackage{xurl}

\usepackage[markup=blue,deletedmarkup=sout, addedmarkup=nf,final]{changes}

\usepackage[figurename=Fig. ]{caption}
\usepackage{subcaption}
\usepackage{wrapfig}
\usepackage[intoc]{nomencl}
\usepackage{etoolbox}
\usepackage{framed} 
\makenomenclature

\renewcommand{\nomgroup}[1]{%
  \item[\bfseries
    \ifstrequal{#1}{S}{Sets}{%
    \ifstrequal{#1}{I}{Indices}{%
    \ifstrequal{#1}{V}{Variables}{%
    \ifstrequal{#1}{P}{Parameters}{}}}}%
  ]
}

\theoremstyle{plain}

\theoremstyle{remark}

\begin{document}
\begin{frontmatter}

\title{Industrial Electrification in the era of data centers: A Bayesian Optimization approach for grid-aware large load allocation}
\author[inst1]{Jiyong Lee}
\author[inst2]{Erhan Kutanoglu}
\author[inst1,inst3]{Michael Baldea}
\author[inst1]{Ilias Mitrai\corref{cor1}}
\ead{imitrai@che.utexas.edu}
\cortext[cor1]{Corresponding author}

\affiliation[inst1]{%
  organization={McKetta Department of Chemical Engineering, The University of Texas at Austin},
  city={Austin},
  postcode={TX 78712},
  country={United States}
}
\affiliation[inst2]{%
  organization={Walker Department of Mechanical Engineering, The University of Texas at Austin},
  city={Austin},
  postcode={TX 78712},
  country={United States}
}
\affiliation[inst3]{%
  organization={Oden Institute for Computational Engineering and Sciences, The University of Texas at Austin},
  city={Austin},
  postcode={TX 78712},
  country={United States}
}

\begin{abstract}
Large loads from industrial electrification and data centers are reshaping the planning and operation of the power grid. Identifying optimal large load siting decisions while accounting for transmission congestion is key to reducing expansion cost and operational risks. In this paper, we propose a leader-follower bilevel optimization framework to identify optimal large load allocation strategies. The leader determines the allocation of large loads, while the followers determine grid expansion cost and transmission utilization. This modeling approach explicitly integrates strategic planning with detailed short-term operational decisions. Moreover, we develop a Bayesian Optimization approach to efficiently solve the bilevel optimization problem by treating the followers as a black box. We use the framework to study large-scale load allocation from electrified oil refineries and data centers on a synthetic power grid that resembles key characteristics of the Texas (ERCOT) system. The results show that these large loads compete for electricity, and under high-load scenarios, data center demand is distributed across the entire grid, avoiding regions with high demand from industrial electrification. 

\end{abstract}


\end{frontmatter}

\section{Introduction} \label{sec: intro}
The rapid growth of large loads from data centers and industrial electrification is reshaping the planning and operation of power systems. Data centers are key infrastructures for deploying artificial intelligence at scale, and are projected to account for 6.7\% to 12\% of the total U.S. electricity demand in 2028 \cite{shehabi20242024}. In parallel, process electrification is emerging as a vital pathway to improve energy efficiency and reduce emissions in the manufacturing sector \cite{baldea2025transforming}. This simultaneous electrification race will require significant grid expansion on short time horizons and will affect grid operations due to the time lag between demand realization and grid expansion \cite{lee2026gridcapacityexpansiondata}. 

An approach to reduce expansion costs while ensuring reliable grid operation is to co-optimize grid expansion and large-load allocation decisions \cite{shao2025stochastic, abdelhady2026optimal}. Existing frameworks rely on representative days to capture grid operation. Although this modeling approach leads to tractable problems, it does not account for extreme-demand days and hours. This limitation is particularly important when considering large loads from data centers and manufacturing.  Data centers can adjust their power consumption over short time scales \cite{dogar2014decentralized}, whereas manufacturing processes are typically subject to much slower dynamics \cite{baldea2025transforming}. The use of representative days for grid expansion and large-load allocation may fail to capture extreme load situations and their effect on grid reliability and stability. 

\begin{figure*}[h]
  \centering
   \includegraphics[scale=0.1]{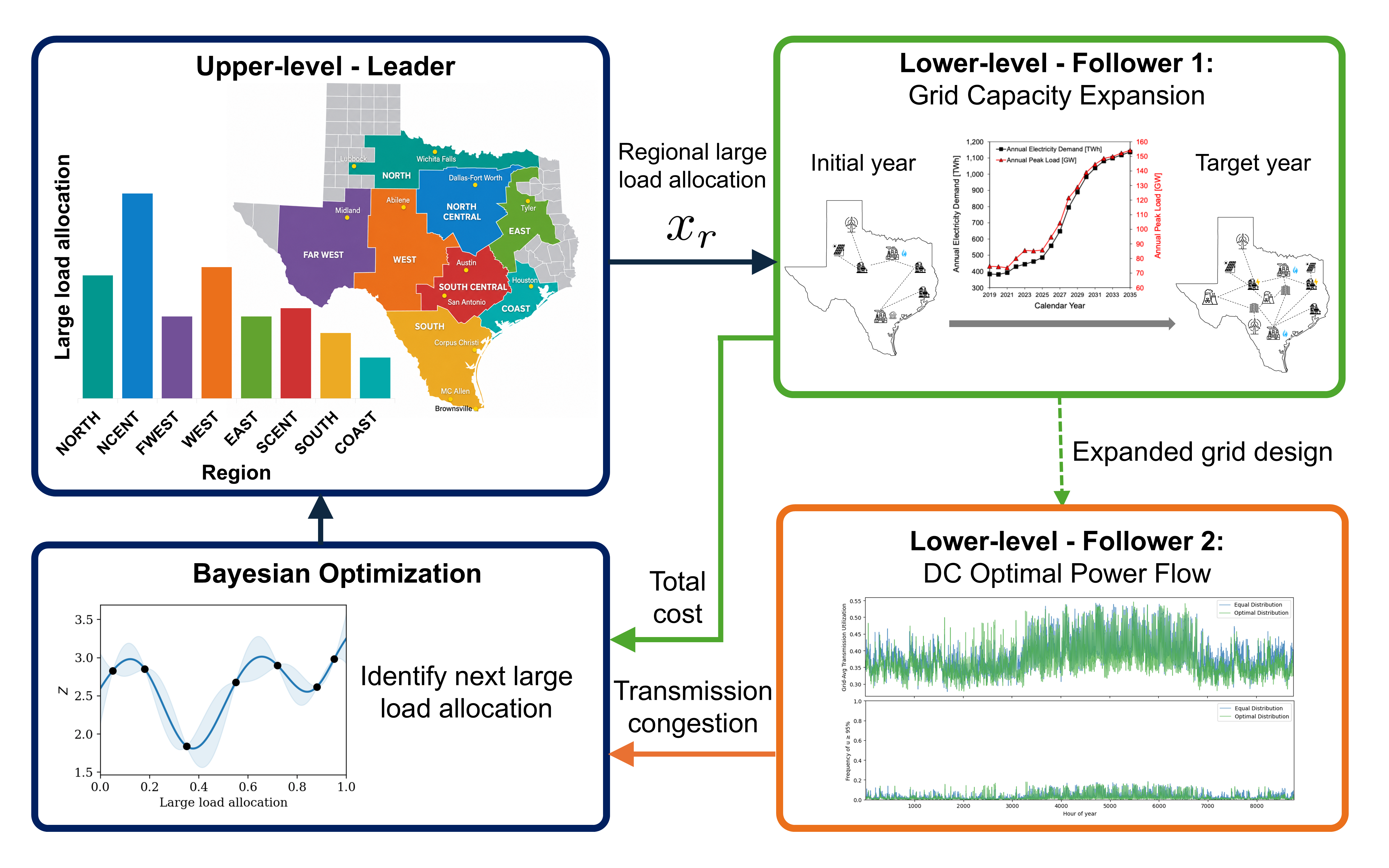}
  \caption{A proposed bilevel optimization formulation and Bayesian Optimization approach for grid-operation-aware large load allocation.}
  \label{fig: bilevel_structure}
\end{figure*}

In this paper, we propose using transmission congestion metrics, computed from detailed optimal power flow calculations, and expansion cost to guide large-load allocation decisions (see Fig.~\ref{fig: bilevel_structure}). First, we formulate the power grid expansion and large load allocation problem as a bilevel optimization problem with one leader, the large load consumers, and two followers determining grid expansion and operation. Second, we develop a Bayesian Optimization approach to efficiently solve the bilevel problem by treating the followers as a black box. As a case study, we use the proposed framework to identify the optimal allocation of data center load for a given electrification target in a synthetic power grid that resembles key characteristics of the Texas power grid (ERCOT) system. The results show that industrial electrification and data centers directly compete for access to electricity. Under high demand, the optimal data center allocation is geographically distributed to reduce expansion cost and improve grid operation.

\section{Bilevel optimization framework} \label{sec: bilevel}
We consider a power grid with $N$ buses, $N_{I}$ power generation technologies, and $N_{l}$ transmission lines. We seek to identify optimal large-load allocation decisions that minimize expansion costs and improve transmission congestion. We assume that the demand from electrified manufacturing is known both in terms of magnitude and spatial distribution. This assumption is driven by the fixed physical location of existing manufacturing facilities. Therefore, we seek to identify the optimal allocation of large loads from data centers across geographical regions to minimize expansion costs and transmission congestion. 
We model this decision-making problem as a leader-follower bilevel optimization problem. 

The leader determines the allocation of the large loads $x$ across regions, which can correspond to individual buses, counties, or weather regions. We define $\mathcal{R}$ as the set of regions over which the large loads must be allocated, and $x_{r} \in [0,1]$ as the fraction of the data center demand allocated in region $r$. The summation of these variables is equal to one enforced via the constraint $\sum_{r \in \mathcal{R}} x_{r} = 1$. We assume that the total data center demand is known and equal to $D^{\rm{DC}}_{\rm{total}}$. 

The problem has two followers: one determining the grid expansion decisions $y_{1}^{*}$ such that grid expansion cost $F_{1}(x,y_{1}^{*})$ is minimized. For a given demand allocation $x$ and optimal grid expansion $y_{1}^{*}$, the second follower computes optimal hourly power dispatch over the entire year $y_{2}^{*}$ which minimizes operating cost $F_{2}(x,y_{1}^{*},y_{2})$. These optimal decisions are used to compute the leader's objective, which accounts for grid expansion $\alpha F_{1}(x,y_{1}^{*})$, with $\alpha$ being a weighting factor, and transmission congestion score $G(y_{1}^{*}, y_{2}^{*})$. Overall, the bilevel problem has the following structure
\begin{equation} \label{eq: boFormulation}
    \begin{aligned}
        \min_{x\in \mathcal{X}} \ \ & \alpha F_{1}(x,y_{1}^{*}) + G(y_{1}^{*}, y_{2}^{*}) \\
        \text{s.t.} \ \ &y_{1}^{*} \in \arg \min_{y_{1} \in \mathcal{Y}_{1}} F_{1}(x, y_{1})\\
        & y_2^* \in \arg \min_{y_{2}\in \mathcal{Y}_{2}} F_{2}(x,y_{1}^{*},y_{2}).
    \end{aligned}
\end{equation}

\subsection{Follower 1 problem - Grid capacity expansion}
For a given large load allocation $x$, first, the power grid must be expanded. We use a single-period capacity expansion model \cite{lee2026cms} that takes as input demand from the residential sector and large loads, the existing grid topology, and associated costs, and identifies optimal expansion investments to minimize capital and operating costs. 

We define $\mathcal{E}$ as the set of counties where the electrified manufacturing facilities are located, $\mathcal{N}$ as the set of buses, $\mathcal{I}$ as the set of power generation technologies, $\mathcal{D}$ as the set of representative days, $\mathcal{H}$ as the set of hours in a day, and $T$ as the target year for which the system is designed. We assume that the grid can expand only existing assets to avoid introducing binary variables and to improve the tractability of the proposed approach.

Investment decisions account for generation capacity technology $i$ at bus $n$, $C^{\rm{gen}}_{ni} \in [0,\bar{C}^{\rm{gen}}]$, transmission between buses, $C^{\rm{trans}}_{nn'}\in [0,\bar{C}^{\rm{trans}}]$, and storage for each bus $C^{\rm{stor}}_\in [0,\bar{C}^{\rm{stor}}]$. The model also includes variables related to power dispatch. Specifically, we define $p^{\rm{gen}}_{nidh}$ as the power generated at bus $n$ using technology $i$ at hour $h$ and representative day $d$, $p^{\rm{trans}}_{nn'dh}$ as the power transmitted between two buses, $p^{\rm{curt,gen}}_{nidh}$ as the curtailment of generation with technology $i$, $p^{\rm{DC}}_{ndh}$ as the nodal load to satisfy data center demand, $p^{\rm{EM}}_{ndh}$ as the nodal load used to satisfy demand from electrified manufacturing,
$p^{\rm{curt,dem}}_{ndh}$ as the curtailment of demand,
$p^{\rm{disch}}_{ndh}$ as the power discharged from storage, and $p^{\rm{charge}}_{ndh}$ as the power charging the storage. The node-level power balance at bus $n$, day $d$, and hour $h$ is
\begin{equation}
\label{eq: powerBalance}
\begin{aligned}
& \sum_{i \in \mathcal{I}(n)} \left(p^{\rm{gen}}_{nidh}- p^{\rm{curt,gen}}_{nidh}\right)
-\sum_{n' \in \mathcal{N}(n)} p^{\rm{trans}}_{nn'dh} + p^{\rm{disch}}_{ndh} - p^{\rm{charge}}_{ndh}\\
 & = D^{\rm{base}}_{ndh} + p^{\rm{DC}}_{ndh} + p^{\rm{EM}}_{ndh} - p^{\rm{curt,dem}}_{ndh}, 
\end{aligned}
\end{equation}
where $p^{\rm{trans}}_{nn'dh}=S_{\rm{base}}(\theta_{ndh}-\theta_{n'dh})/X_{nn'}$ is the power transmitted, $D^{\rm{base}}_{ndh}$ is the base load at bus $n$, day $d$, and hour $h$, $\theta_{ntdh} \in [-\pi,\pi]$ is the voltage angle in radians, the system base power $S_{\rm{base}}$ is 100 MVA, and $X_{nn'}$ is the line reactance. The data center and electrified manufacturing loads are satisfied via the following equations
\begin{equation}
    \sum_{n \in \mathcal{N}_{r}} p^{\rm{DC}}_{ndh} = D^{\rm{DC}}_{rdh} \quad \forall r \in \mathcal{R}, d \in \mathcal{D}, h \in \mathcal{H},
    \label{eq: county-level DC load satisfaction}
\end{equation}
\begin{equation}
    \sum_{n \in \mathcal{N}_{e}} p^{\rm{EM}}_{ndh} = D^{\rm{EM}}_{edh} \quad \forall e \in \mathcal{E}, d \in \mathcal{D}, h \in \mathcal{H},
    \label{eq: county-level EOR load satisfaction}
\end{equation}
where $D^{\rm{DC}}_{rdh}$ and $D^{\rm{EM}}_{edh}$ are data center and electrified manufacturing load applied on region $r$ and county $e$, respectively, on representative day $d$, hour $h$. The data center demand for each region $r$ is equal to
\begin{equation}
    D^{\rm{DC}}_{rdh} = x_{r} D^{\rm{DC}}_{\rm{total}}/(|\mathcal{D}| |\mathcal{H}|) \ \ \forall r \in \mathcal{R},
\end{equation}
and is determined by the leader. We assume that total data center load is evenly distributed across all days and hours of the year. However, the capacity expansion model determines how to allocate demand across buses within a region. 

The power generated, transmitted, and stored for each hour $h$ and representative day $d$ is bounded by the installed capacity, i.e., $p^{\rm{gen}}_{nidh} \leq C^{gen}_{ni}$, $|p^{\rm{trans}}_{nn'dh}| \leq C^{\rm{trans}}_{nn'}$, $p^{\rm{charge}}_{ndh} \leq C^{\rm{stor}}_{n}$, $p^{\rm{dich}}_{ndh} \leq C^{\rm{stor}}_{n}$. Finally, we include constraints related to storage, which we omit due to space limitations (see \cite{lee2026gridcapacityexpansiondata}). The objective function accounts for capital and operating costs and is linear in the decision variables. We omit the presentation here due to space limitations (see \cite{lee2026gridcapacityexpansiondata} for a detailed description). The optimal solution of this problem is denoted as $y_{1}^{*}$.

\subsection{Follower 2 problem - DC Optimal power flow}
The second follower, for a fixed data center demand allocation and grid expansion, solves a DC-optimal power flow model to compute the hourly power dispatch decisions for the entire year. This model is similar to the capacity expansion model presented above, but the capacities are fixed and the decision horizon is the full year with 365 days, i.e., we do not use representative days, and the objective only accounts for the operating cost. We denote the optimal power transmission decisions from this model as $\hat{p}^{\rm{trans}}_{nn'dh}$ with $d \in \{1,\dots, 365\}$, and the optimal solution vector as $y_{2}^{*}$.

\subsection{Objective of the Leader}
The leader's objective is the sum of the grid expansion cost and the transmission congestion score. The capacity expansion cost is equal to the optimal objective function value of the first follower, denoted as $F_{1}(x,y_{1}^{*})$. To quantify the utilization rate of each transmission line at each operational time, the transmission utilization rate is defined as 
\begin{equation} \label{eq: trans_utilization}
u^{\rm{trans}}_{nn'dh}=|\hat{p}^{\rm{trans}}_{nn'dh}|/C^{\rm{trans}}_{nn'},
\end{equation}
where $\hat{p}^{\rm{trans}}_{nn'dh}$ is obtained from the optimal solution of the second follower and $C^{\rm{trans}}_{nn'}$ is obtained from the optimal solution of the first follower. Note that the transmission lines are assumed to be bidirectional, so we calculate the absolute magnitude of power flow. The average grid-wide transmission utilization rate is equal to 
\begin{equation} \label{eq: grid_avgUtilization}
    \bar{u}^{\rm{trans}}_{\rm{grid}}=
    \frac{1}{N_{l} \times 365 \times |\mathcal{H}|}
    \sum_{(n,n')\in\mathcal{L}}
    \sum_{d=1}^{365}
    \sum_{h\in\mathcal{H}}
    u^{\rm{trans}}_{nn'dh},
\end{equation}
where the $\mathcal{L}$ is the set of transmission lines. This metric measures the network's average grid-wide transmission load during the year. We also compute the frequency with which the power transmitted between two buses over the capacity is greater than or equal to $70\%$ and $95\%$ as follows
\begin{equation} \label{eq: frequency70}
    \rho^{70\%}_{\rm{grid}}=
    \frac{1}{N_{l} \times 365 \times |\mathcal{H}|}
    \sum_{(n,n')\in\mathcal{L}}
    \sum_{d\in\mathcal{D}}
    \sum_{h\in\mathcal{H}}
    \mathbb{I}\left(u^{\rm{trans}}_{nn'dh}\geq 0.70\right),
\end{equation}
\begin{equation} \label{eq: frequency95}
    \rho^{95\%}_{\rm{grid}}=
    \frac{1}{N_{l} \times 365 \times |\mathcal{H}|}
    \sum_{(n,n')\in\mathcal{L}}
    \sum_{d\in\mathcal{D}}
    \sum_{h\in\mathcal{H}}
    \mathbb{I}\left(u^{\rm{trans}}_{nn'dh}\geq 0.95\right),
\end{equation}
where $\mathbb{I}(\cdot)$ is an indicator function that equals one when the condition is true and zero otherwise. Finally, we combine these three utilization and congestion-related metrics into a transmission congestion score, which is defined as
\begin{equation} \label{eq: transCongestionScore}
G=\lambda_1 \bar{u}^{\rm{trans}}_{\rm{grid}} + \lambda_2 \rho^{70\%}_{\rm{grid}} + \lambda_3 \rho^{95\%}_{\rm{grid}},
\end{equation}
where $\lambda_1,\lambda_2$, and $\lambda_3$ are weights. Overall, the objective of the leader is equal to $\alpha F_{1} (x,y_{1}^{*}) + G(y_{1}^{*}, y_{2}^{*})$. The resulting problem is a nonlinear discrete bilevel optimization problem, due to constraints \ref{eq: grid_avgUtilization}, \ref{eq: frequency70}, \ref{eq: frequency95}, with linear lower-level problems. 

\section{Bayesian Optimization solution approach}\label{sec: bo}
The solution of the bilevel problem presented above is challenging due to the presence of the indicator functions, which make the objective nonsmooth. Although the problem admits an exact reformulation as a single- level problem using the KKT optimality conditions for the linear lower-level problems, the resulting problem will be a large-scale discrete optimization problem. To solve the bilevel problem, we exploit its underlying structure by noting that if the large-load allocation decisions are fixed, the objective can be evaluated after solving the lower-level problems and computing the transmission congestion score. Motivated by this structure, we treat the lower-level problems as a black box and use Bayesian Optimization (BO) to identify the optimal allocation of the large loads. 

BO uses a Gaussian Process (GP) $m$ to approximate the objective function $\alpha F_{1}(x,y_{1}^{*})+G(y_{1}^{*}, y_{2}^{*})$. A GP is a distribution over functions defined as 
$m(x) \sim \mathcal{GP}(\mu(\cdot), k(\cdot,\cdot))$, where $\mu(\cdot)$ is the mean function, $k(\cdot,\cdot)$ is the covariance kernel, and $x = \{x_{r}\}_{r \in \mathcal{R}}$. To navigate the large load allocation design space, we use the logarithmic expected improvement as the acquisition function. The proposed approach alternates between suggesting new large load allocations, $x$, and solving the capacity expansion and DC-OPF models to obtain $y_{1}^{*}$ and $y_{2}^{*}$ and compute $\alpha F_{1}(x,y_{1}^{*})+G(y_{1}^{*},y_{2}^{*})$. The BO approach is implemented in BoTorch \cite{balandat2020botorchframeworkefficientmontecarlo}, the capacity expansion and DC-OPF models are implemented in Pyomo \cite{hart2017pyomo}.
\begin{table}[t]
  \caption{Load scenarios considered in the study. DC refers to Data Center; EOR refers to Electrified Oil Refining. The base case scenario is marked with an asterisk (*).}
  \label{tab:demand_peak_scenarios}
  \centering
  \renewcommand{\arraystretch}{1.08}
  \footnotesize
  \begin{tabular}{@{} l r r r r r @{}}
    \toprule
    \multicolumn{2}{c}{\textbf{Load Scenario}} &
    \multicolumn{4}{c}{\textbf{Annual Electricity Demand [TWh]}} \\
    \cmidrule(lr){1-2}
    \cmidrule(lr){3-6}
    \textbf{DC} & \textbf{EOR} &
    \textbf{Base} & \textbf{DC} & \textbf{EOR} & \textbf{Total} \\
    \midrule
    low       & 25\%  & 846.99 &  95.38 &  42.82 &  985.19 \\
    moderate  & 25\%  & 846.99 & 190.75 &  42.82 & 1080.57 \\
    high      & 25\%  & 846.99 & 381.51 &  42.82 & 1271.32 \\
    low       & 50\%  & 846.99 &  95.38 &  85.65 & 1028.01 \\
    moderate* & 50\%* & 846.99 & 190.75 &  85.65 & 1123.39 \\
    high      & 50\%  & 846.99 & 381.51 &  85.65 & 1314.14 \\
    low       & 100\% & 846.99 &  95.38 & 171.29 & 1113.66 \\
    moderate  & 100\% & 846.99 & 190.75 & 171.29 & 1209.04 \\
    high      & 100\% & 846.99 & 381.51 & 171.29 & 1399.79 \\
    \bottomrule
  \end{tabular}
\end{table}

\begin{figure*}[t] 
  \centering
  \includegraphics[scale=0.5]{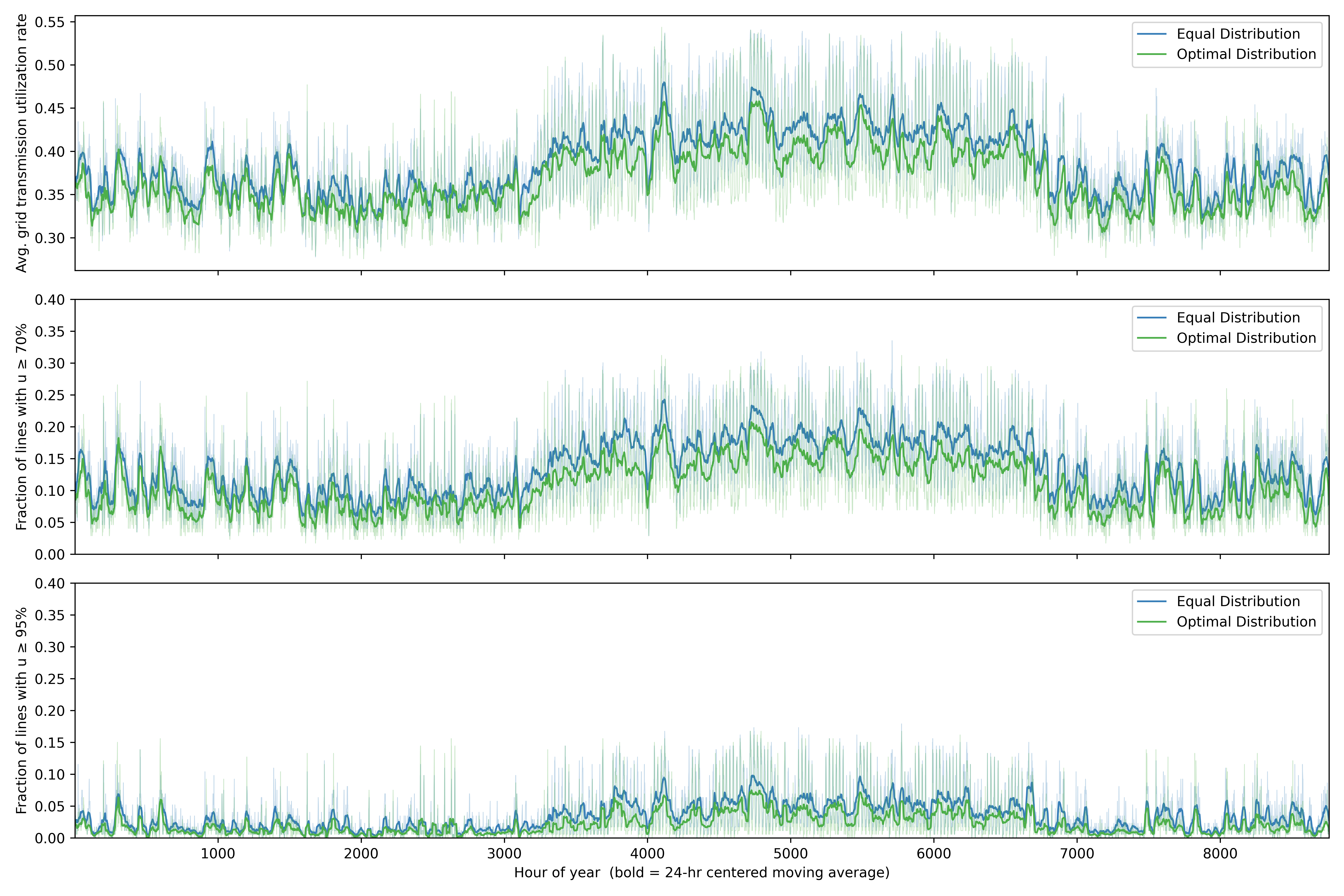}
  \caption{Comparison of transmission utilization and frequency of transmission congestion for equal and optimal distribution of data center load obtained by solving the full-year DC-OPF model. Faint line graphs represent hourly raw data, and bold line graphs represent daily averaged trends.}
  \label{fig: baseCase_fullYearDC-OPF}
\end{figure*}

\section{Case study} \label{sec: case}
In this section, we use the proposed approach to identify optimal allocations of data center loads for a given reference grid and industrial electrification strategy. The grid, based on the Texas ERCOT network, is adapted from \cite{lee2026gridcapacityexpansiondata}, \cite{lu2025synthetic}, and has 123 buses, 151 generators across six generation technologies (coal, natural gas, nuclear, wind, solar, and hydro), 173 bidirectional transmission lines (345 kV), and initially zero storage capacity. Other operational and economic parameters can be found in \cite{lee2026cms, lee2026gridcapacityexpansiondata}. We consider 2031 the target year ($T=2031$) and evaluate grid expansion and operation under the specified load growth scenario. Given that we do not explicitly consider the construction time of grid assets, which can be an important factor in guiding investment policy~\cite{lee2026gridcapacityexpansiondata}, we impose an upper limit of 1GW on newly invested generation, transmission, and storage capacity, i.e., $\bar{C}^{\rm{gen}}=\bar{C}^{\rm{trans}}=\bar{C}^{\rm{stor}}=1\rm{GW}$. This upper bound acts as a proxy for construction time limitations by preventing unrealistically large capacity investments within the single grid capacity expansion problem. The weight constants for the transmission congestion score in Eq.~\ref{eq: transCongestionScore} are equal to $\lambda_1=1$, $\lambda_2=5$, and $\lambda_3=10$, and $\alpha=10^{-11}$. For BO and each demand scenario, we use 16 initial demand allocations and allow 84 iterations (100 evaluations in total). The first point is fixed as the uniform allocation, i.e., $x_{r}=1/|\mathcal{R}|$, and the remaining 15 initial points are sampled from a Dirichlet distribution over the feasible space of allocations. 

The single-year capacity expansion problem has approximately $203 \times 10^{3}$ constraints and $128 \times 10^{3}$ variables, while the full-year DC-OPF model has approximately $11.8 \times 10^{6}$ constraints and $9.3 \times 10^{6}$ variables. All instances are solved using Gurobi v13.0.2 with the barrier method on an Intel Xeon CPU Max 9480.  For the base case scenario (DC: moderate, EOR: 50\%), the average time per BO iteration is 715 seconds, and the total solution time for 100 iterations is 16 hours.

\subsection{Load growth scenarios} \label{sec:loadGrowth}
The grid must satisfy three types of loads: base, data center, and electrified oil refining (EM) demand, which are computed as follows:
\begin{equation}
\begin{aligned}
D^{\rm{base}}_{ndh} & = D^{\rm{base},0}_{ndh} + \Delta D^{\rm{base}}\\
D^{\rm{DC}}_{rdh} & = x^{DC}_{r} D^{\rm{DC}}_{\rm{total}}\\
D^{\rm{EM}}_{edh} & = \psi^{EM}_{e} (\phi/\eta_{elec}) Q^{\rm{M}}.
\end{aligned}
\label{eq: demand}
\end{equation}
For the base load, $D^{\rm{base},0}_{ndh}$ is the initial base load profile obtained from \cite{lu2025synthetic} at bus $n$ day $d$ and hour $h$, $\Delta D^{\rm{base}}$ is the increase in base demand between the current year $t_{0}$ and target year $T$. This increase is computed using the initial and projected annual base demand. We assume that the increase in base demand is distributed equally across all buses, days, and hours.

For data center load, the total demand, $D^{\rm{DC}}_{\rm{total}}$, is calculated using the annual peak load projection with a load factor of $90\%$ \cite{lee2026gridcapacityexpansiondata}. Finally, for the demand from electrified manufacturing, we use publicly available data to obtain the heat demand from oil refineries per county in Texas, $Q^{\rm{M}}$, for conventional boilers (100 - 400 $^{o}C$) and process heating ($\geq 400 ^{o}C$) \cite{mcmillan2018industrial}. We define $\phi$ as the fraction of the heat demand that is electrified in the target year. We also define $\eta_{elec}$ as the Joule-heating efficiency, which is assumed to be 97\%. For both large loads, we assume that the hourly load profile is constant throughout the year. A detailed explanation of modeling the base and large load electricity demands can be found at \cite{lee2026gridcapacityexpansiondata}.

Based on these available data, we consider nine load growth scenarios. For data center load, we use ERCOT's 2031 projection \cite{ERCOT2025LTLF} as the \textit{moderate} scenario. We then define the \textit{low} and \textit{high} data center load scenarios with 50\% and 200\% of the \textit{moderate} scenario, respectively. For the electrified manufacturing, we consider three scenarios with electrification ratio $(\phi)$ equal to 25\%, 50\%, and 100\%. For every scenario, we assume the base load remains constant at the projected 2031 value from the ERCOT report \cite{ERCOT2025LTLF}. The resulting nine load scenarios are presented in Table~\ref{tab:demand_peak_scenarios}. We define the \textit{base case scenario} as the combination of the \textit{moderate} data center load scenario and the \textit{50\%} electrification scenario. Code implementation can be found on
\href{https://github.com/PSE-Lab/Bilevel-Optimization-Framework-for-Grid-aware-Large-Load-Allocation.git}{GitHub} \cite{lee2026bilevelGit}.

\section{Results} \label{sec: results}
We first analyze the base case scenario and compare the results between equal and optimal data center load allocation. First, we find that BO improves the leader's objective primarily by reducing the transmission congestion score ($G$) rather than the expansion and operation cost ($\alpha F_1$). From Table~\ref{tab:base_case_bo_results}, the optimal allocation does not assign data center demand to the coast and west weather regions, decreasing the leader's objective from 3.0083 to 2.7428. The capacity expansion cost remains the same for both allocations, i.e., the new installed capacity is the same. The transmission congestion ($G$) score reduces from 1.3914 to 1.1259. Specifically, the average grid-level transmission utilization rate decreases by 1.82\% point (from 38.53\% to 36.71\%), the frequency of line-hour exceeding 70\% decreases by 2.59\% point (from 13.40\% to 10.81\%), and the frequency of line-hour exceeding 95\% decreases by 1.18\% point (from 3.36\% to 2.18\%). 

\begin{table}
\caption{Base case scenario results: data center load allocation to different weather zones and Bayesian optimization observations.}
\centering
\begin{tabular}{l rr}
\toprule
\textbf{} & \textbf{Equal} & \textbf{Optimal} \\
\midrule
$x^{DC}_{COAST}$  & 12.50\% &  0.00\% \\
$x^{DC}_{EAST}$   & 12.50\% & 19.12\% \\
$x^{DC}_{FWEST}$  & 12.50\% &  7.22\% \\
$x^{DC}_{NCENT}$  & 12.50\% & 22.87\% \\
$x^{DC}_{NORTH}$  & 12.50\% & 19.33\% \\
$x^{DC}_{SCENT}$  & 12.50\% & 21.41\% \\
$x^{DC}_{SOUTH}$  & 12.50\% & 10.05\% \\
$x^{DC}_{WEST}$   & 12.50\% &  0.00\% \\
\midrule
$\alpha F_{1}$        & 1.6169  &  1.6168 \\
$G$                   & 1.3914  &  1.1259 \\
$Z=\alpha F_{1}+G$    & 3.0083  &  2.7428 \\
\bottomrule
\end{tabular}
\label{tab:base_case_bo_results}
\end{table}
The results also show that the optimal allocation leads to lower grid-wide transmission utilization and less congestion (see Fig.~\ref{fig: baseCase_fullYearDC-OPF}). For both allocation strategies, the average grid-wide transmission utilization rates ($\bar{u}^{\rm{trans}}$) range from 28\% to 54\%. However, the average value of the equal distribution is 38.53\%, while for the optimal distribution, it is 36.71\%. Frequencies of line-hour congestion at 70\% and 95\% utilization are both reduced under the optimal distribution allocation. The largest relative improvement is in the severe congestion metric with 95\% of utilization ($\rho^{95\%}_{\rm{grid}}$). The average and maximum $\rho^{95\%}_{\rm{grid}}$ values of equal distribution are 3.36\% and 17.92\%, while those of optimal distribution are 2.18\% and 16.76\%. 
\begin{figure}
  \centering
   \includegraphics[width=\columnwidth]{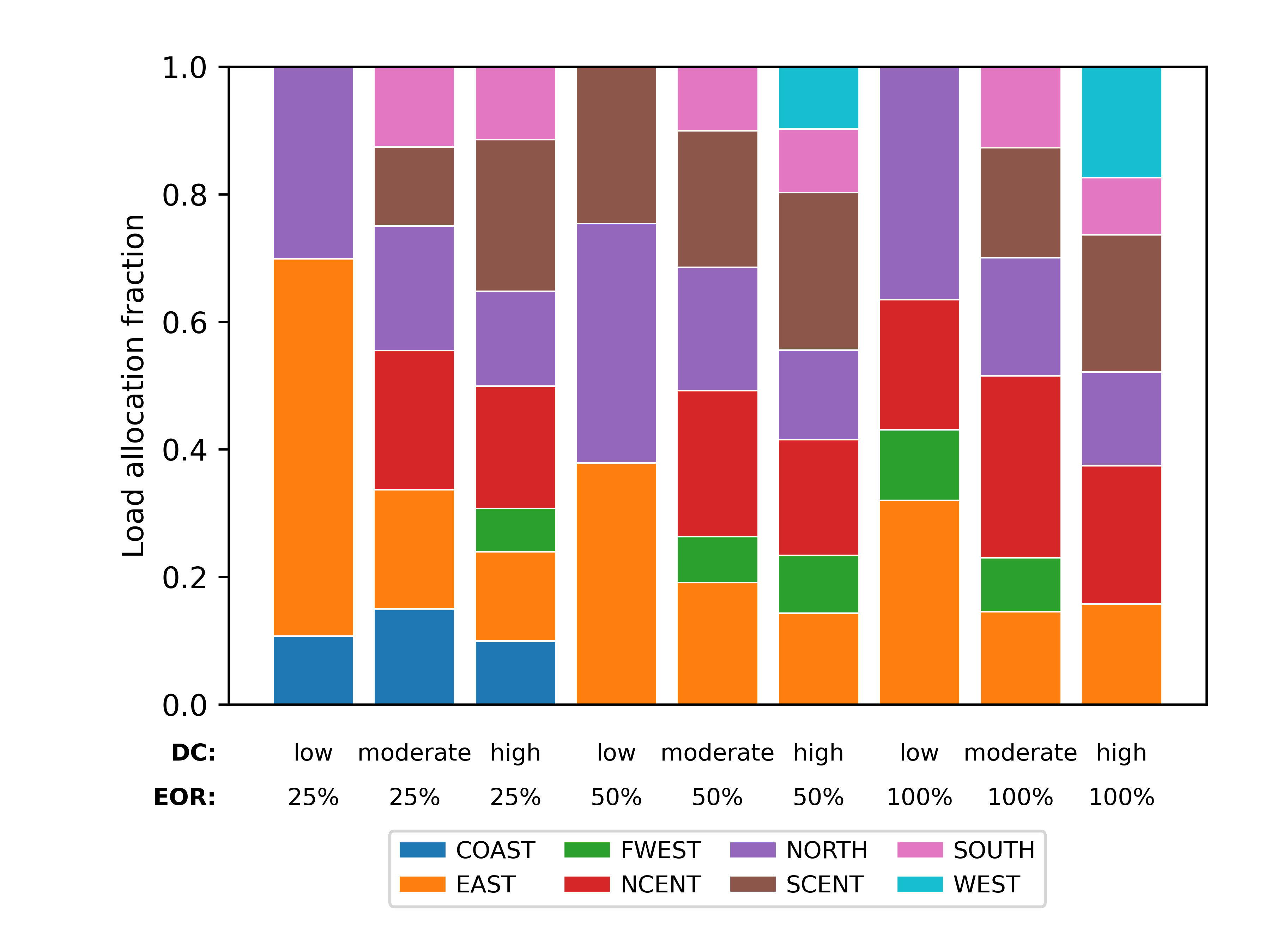}
  \caption{Optimal data center load allocation across weather regions for multiple load scenarios.}
  \label{fig: MultiLoad_Allocation}
\end{figure}

Finally, we analyze the allocation of data center loads for the different load scenarios presented in Table~\ref{tab:demand_peak_scenarios}. From Fig.~\ref{fig: MultiLoad_Allocation}, we observe that for a fixed electrification ratio, an increase in data center demand leads to a more spatially distributed allocation. Under low data center and 25\% electrified loads, data center load is assigned to the east, north, and coast regions. However, as the data center load increases, the allocation becomes more distributed across north, ncent, south, scent, fwest, east, and coast. This suggests that concentrating data center load in limited regions becomes less favorable as the total large load magnitude increases.

Moreover, the electrification ratio of oil refineries directly affects the allocation of data center loads. Specifically, we find that data center load allocation to the coast region becomes negligible once the electrification ratio exceeds 25\%. Since the oil refining industry in Texas is mostly concentrated in the coast region, the optimal allocation avoids placing data center loads in this region to reduce capacity expansion costs and improve transmission congestion. This result suggests that the capacity expansion model with representative days may miss full-year operational grid stress patterns. The full-year DC-OPF model captures these temporal grid congestion effects and informs the allocation of large loads.

\section{Conclusions} \label{sec: conclusion}
In this paper, we consider the design of large load allocation policies that account for detailed transmission-congestion-related performance and capacity expansion costs. We propose a bilevel optimization framework to identify optimal grid expansion and allocation of large loads such that grid expansion cost and transmission congestion are optimized. We develop a Bayesian Optimization approach to efficiently solve the resulting bilevel problem. An application to a synthetic power grid under large loads from data centers and electrified oil refining shows that the proposed approach identifies large load allocations that reduce grid congestion without increasing expansion costs. 

\section{Acknowledgements}
Financial support from the Energy Institute at The University of Texas at Austin is gratefully acknowledged. 

\bibliographystyle{elsarticle-num}
\bibliography{refs}

@article{shehabi20242024,
  author    = {Shehabi, Arman and Smith, Sherrill J. and Hubbard, Aubrey and 
               Newkirk, Austin and Lei, Nuoa and Siddik, Md Arman Bilal and 
               Holecek, Bryan and Koomey, Jonathan and Masanet, Eric and 
               Sartor, Dale},
  title     = {2024 {United States} Data Center Energy Usage Report},
  year      = {2024},
  journal = {Lawrence Berkeley National Laboratory},
  address   = {Berkeley, California},
  note      = {LBNL-2001637}
}

@article{baldea2025transforming,
  title={Transforming the Process Industries through Electrification: Challenges and Opportunities},
  author={Baldea, Michael and Endler, Elizabeth E and Hale, Elaine and Maravelias, Christos T and Barolo, Massimiliano and Harjunkoski, Iiro and Mercangoz, Mehmet and Shah, Sirish L and Soroush, Masoud and Young, Brent R and others},
  journal={Ind. Eng. Chem. Res.},
  volume={64},
  number={34},
  pages={16466--16478},
  year={2025},
  publisher={ACS Publications}
}

@article{lee2026gridcapacityexpansiondata,
  title={Grid Capacity Expansion under Data Centers and Electrified Manufacturing Large Loads},
  author={Lee, Jiyong and Agustin, Melody and Langsdorf, Joanne and Kutanoglu, Erhan and Baldea, Michael and Mitrai, Ilias},
  journal={arXiv preprint arXiv:2605.29053},
  year={2026}
}

@article{shao2025stochastic,
  title={Stochastic long-term joint decarbonization planning for power systems and data centers: A case study in PJM},
  author={Shao, Zhentong and Yu, Nanpeng and Wong, Daniel},
  journal={International Journal of Electrical Power \& Energy Systems},
  volume={173},
  pages={111377},
  year={2025},
  publisher={Elsevier}
}

@article{abdelhady2026optimal,
  title={Optimal energy portfolio investment strategies for data centers under deep market uncertainty},
  author={Abdelhady, Mohamed and Iakovou, Eleftherios and Pistikopoulos, Efstratios N},
  journal={Applied Energy},
  volume={410},
  pages={127510},
  year={2026},
  publisher={Elsevier}
}

@article{dogar2014decentralized,
  title={Decentralized task-aware scheduling for data center networks},
  author={Dogar, Fahad R and Karagiannis, Thomas and Ballani, Hitesh and Rowstron, Antony},
  journal={ACM SIGCOMM Computer Communication Review},
  volume={44},
  number={4},
  pages={431--442},
  year={2014},
  publisher={ACM New York, NY, USA}
}

@article{lee2026cms,
    title={Effect of data center and electrified manufacturing demand on power grid
    expansion},    
    author={Lee, Jiyong and Kutanoglu, Erhan and Baldea, Michael and Mitrai, Ilias},
    journal={Procedia CIRP},
    year={2026},
    note={In press}
}

@book{hart2017pyomo,
  title={Pyomo-optimization modeling in python},
  author={Hart, William E and Laird, Carl D and Watson, Jean-Paul and Woodruff, David L and Hackebeil, Gabriel A and Nicholson, Bethany L and Siirola, John D and others},
  volume={67},
  year={2017},
  publisher={Springer}
}

@article{balandat2020botorchframeworkefficientmontecarlo,
  title={BoTorch: A framework for efficient Monte-Carlo Bayesian optimization},
  author={Balandat, Maximilian and Karrer, Brian and Jiang, Daniel and Daulton, Samuel and Letham, Ben and Wilson, Andrew G and Bakshy, Eytan},
  journal={Adv. Neural Inf. Process. Syst.},
  volume={33},
  pages={21524--21538},
  year={2020}
}

@article{lu2025synthetic,
  title={A Synthetic Texas Power System with Time-Series Weather-Dependent Spatiotemporal Profiles},
  author={Lu, Jin and Li, Xingpeng and Li, Hongyi and Chegini, Taher and Gamarra, Carlos and Yang, YC Ethan and Cook, Margaret and Dillingham, Gavin},
  journal={Sustainable Energy, Grids and Networks},
  pages={101774},
  year={2025},
  publisher={Elsevier}
}

@techreport{mcmillan2018industrial,
    title={Industrial process heat demand characterization},
  author={McMillan, Colin and Ruth, Mark},
  year={2018},
  institution={National Renewable Energy Laboratory},
  type         = {Technical Report}
    
}

@techreport{ERCOT2025LTLF,
  author       = {{Electric Reliability Council of Texas (ERCOT)}},
  title        = {2025 ERCOT System Planning: Long-Term Hourly Peak Demand and Energy Forecast},
  year         = {2025},
  month        = apr,
  type         = {Technical Report}
}

@software{lee2026bilevelGit,
  author  = {Lee, Jiyong and Kutanoglu, Erhan and Baldea, Michael and Mitrai, Ilias},
  title   = {Bilevel Optimization Framework for Grid-aware Large Load Allocation},
  year    = {2026},
  url     = {https://github.com/PSE-Lab/Bilevel-Optimization-Framework-for-Grid-aware-Large-Load-Allocation},
  note    = {GitHub repository},
}

\end{document}